\documentstyle[twocolumn,aps,prl,floats,epsf,multirow]{revtex}

\begin{document}

\def\pt{$ p_{t}$ } \def\xf{$ x_F$ } \def\rt{$ \sqrt\tau$ }
\def\u{$\Upsilon$ } \def\JP{$\psi'$ } \def\J{$J/\psi$ }
\def\up{$\Upsilon^{\prime}$ } \def\upp{$\Upsilon^{\prime\prime}$ }
\def\D{$^2H$ }

\draft

\wideabs{

\title{Energy loss of fast quarks in nuclei}                                
\author{
M.B.~Johnson$^a$,
B.Z.~Kopeliovich$^b$,
I.K.~Potashnikova$^b$, \\
and \\
P.L.~McGaughey$^a$,
J.M.~Moss$^a$,
J.C.~Peng$^a$,
G.T.~Garvey$^a$,
M.J.~Leitch$^a$,
M.R.~Adams$^c$,
D.M.~Alde$^a$,
H.W.~Baer$^{a,f}$\cite{byline1},
M.L.~Barlett$^d$,
C.N.~Brown$^e$,
W.E.~Cooper$^e$,
T.A.~Carey$^a$,
G.~Danner$^{a,f}$,
G.W.~Hoffmann$^d$,
Y.B.~Hsiung$^e$,
D.M.~Kaplan$^g$\cite{byline2},
A.~Klein$^a$\cite{byline3},
C.~Lee$^a$,
J.W.~Lillberg$^a$,
R.L.~McCarthy$^h$,
C.S.~Mishra$^a$\cite{byline4},
M.J.~Wang$^f$\cite{byline5}
\\ \vspace*{5pt}
(FNAL E772 Collaboration)\\ \vspace*{9pt}
}
\address{
$^a$Los Alamos National Laboratory, Los Alamos, NM 87545\\
$^b$Max-Plank-Institut f\"ur Kernphysik, 69029 Heidelberg, Germany\\
$^c$University of Illinois, Chicago, IL 60510\\
$^d$Univerity of Texas, Austin, TX  78712\\
$^e$Fermi National Accelerator Laboratory, Batavia, IL 60510\\
$^f$Case Western Reserve University, Cleveland, OH 44106\\
$^g$Northern Illinois University, DeKalb, IL 60115\\
$^h$State University of New York at Stony Brook, Stony Brook, NY 11794\\
}
\date{\today}

\maketitle

\begin{abstract}

We report an analysis of the nuclear dependence of the yield of
Drell-Yan dimuons from the 800 GeV/c proton bombardment of 
$^2H$, C, Ca, Fe, and W targets.  Employing a new formulation of the
Drell-Yan process in the rest frame of the nucleus, this analysis
examines the effect of initial-state energy loss and shadowing 
on the nuclear-dependence 
ratios versus the incident proton's momentum fraction and dimuon 
effective mass.
The resulting energy loss per unit path length is 
$-dE/dz = 2.32 \pm 0.52\pm 0.5$ GeV/fm.
This is the first observation of a nonzero energy loss of partons 
traveling in nuclear environment.

\end{abstract}

} 


For many years it has been suggested that quark energy loss might give 
rise to a nuclear dependence\cite{bbl,bodwin,kn1,gm} of the cross section 
of Drell-Yan\cite{drellyan} (DY) production. When a proton enters a 
nucleus the first (soft) inelastic collision liberates a quark, which then 
loses energy via hadronization (due to confinement) and interaction in 
the nuclear medium.
A lepton pair created from a subsequent interaction then has reduced 
energy compared with the DY process on a free nucleon. The goal of the 
present analysis is to search for this effect in the nuclear dependence 
of the DY process.

Fermilab E772 made a precise measurement of the nuclear dependence of
the DY process using 800 GeV/c protons.  
The experimental details of E772 have been described
previously\cite{aldedy,aldeups,mcgaughey}.  Briefly we indicate those germane 
to the present discussion.  Muon pairs were recorded from targets of $^2$H,
C, Ca, Fe, and W, in the mass range $M \geq 4$ GeV/$c^2$. 
Excluding the \u resonance region, $9\leq M\leq 11$ GeV/$c^2$, we 
reconstruct
$2.5\times10^5$ DY dimuons. The spectrometer acceptance for this 
subset of the data had transverse momentum coverage out to 3.5 GeV/c.  
Since E772 was designed to make a precision comparison of the yields of dimuons from 
the heavy targets to that from $^2$H, relative target-to-target normalization 
errors were kept to $\leq$ 2\%.

The parton model description of high-energy processes is reference-frame 
dependent. Because energy loss is most commonly described in the
rest frame of the nucleus, it is best to adopt this frame for the
description of the DY process as well.  In the target rest frame the DY 
process for proton-nucleon collisions 
is treated as bremsstrahlung~\cite{hir}: An incident quark with momentum 
fraction $x_q$ emits a virtual photon that carries a fraction 
$x_1^q =x_1/x_q$ of the quark momentum. 
The inclusive cross section for the production of lepton pairs with
momentum fraction $x_1$ is given by
\begin{eqnarray}
{d\sigma_{DY}^{pN}(M^2)\over dx_1}=\int_{x_1}^1
dx_qF_q^p(x_q){d\sigma_{DY}^{qN}(x_1^q,M^2)\over dx_1^q}, \label{eq:signn}
\end{eqnarray}
where $F_q^p(x_q)$ is the quark distribution function of
the proton and $d\sigma_{DY}^{qN}(x_1^q,M^2)/dx_1^q$ is the quark-nucleon
differential cross section for lepton-pair production\cite{hir,bhq,kst1}.

\begin{figure}[thb]
  \begin{center}
    \mbox{\epsfxsize=2.5in\epsffile{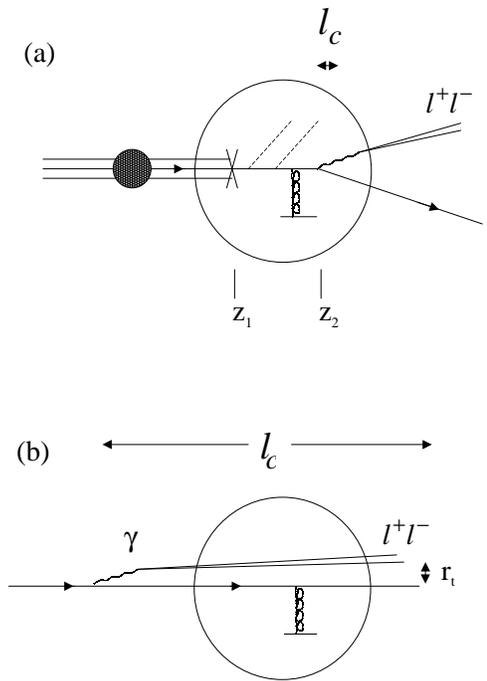}} 
    \vspace*{+9pt}                                                      
  \end{center}                                                           
 
\caption{Schematic representation, in the nuclear rest frame, of two processes 
producing a nuclear dependence of the DY cross section. (a): Energy 
loss. A proton 
entering a nucleus undergoes its first inelastic collision at point 
$z_1$, liberating a quark. The quark propagates, losing energy to 
hadronization (dashed lines), to point $z_2$ where it undergoes a hard 
(DY) interaction, producing a virtual photon that decays into a lepton 
pair. (b): Shadowing. A fast quark undergoes a virtual 
fluctuation into a photon and quark. The quark propagates into the 
nucleus, liberating the fluctuation. Since the coherence length is large, 
the entire nucleus participates as a single entity.}
  \label{models}
\end{figure}

Nuclear effects modify this in two important ways.
The first is the possibility of quark energy loss 
in the nuclear medium -- 
the main subject of this manuscript. The second is shadowing, a phenomenon 
well known from nuclear dependence studies\cite{geesaman} of a closely related 
process, deeply-inelastic lepton scattering (DIS). Energy loss and 
shadowing are shown pictorially in Fig.~\ref{models}. Since the two 
processes produce apparently similar effects in 
proton-nucleus collisions, it is necessary to adopt a consistent analysis 
where both are considered on the same footing. The framework for 
accomplishing this is detailed in the following paragraphs, first for energy 
loss, then for shadowing.

Consider a proton entering a nucleus (Fig.~\ref{models}).    
The first inelastic interaction, at point $z_1$, removes the 
coherence among the soft projectile partons, which then move apart losing 
energy as they would in the vacuum. A quark continues to propagate 
arriving at point $z_2$ where a DY interaction takes place with diminished
energy $\tilde x_q\,E_p=x_q\,E_p-\Delta E$, where $\Delta E$ is the energy loss to
be measured and $E_p$ is the energy of the proton beam.  Correspondingly, one has
$\tilde x^q_1=
x_1/(x_q-\Delta E/E_p)$.  Assuming
the rate of energy loss is constant, $\Delta E \propto L$, where $L=z_2-z_1$.
Due to energy loss, the ratio of p-A to p-N cross sections versus $x_1$ is 
\begin{eqnarray} R_{A/N}^{\Delta E}(x_1,M^2)=
{\int_{x_1+\Delta E/E_p}^1 dx_q\, F_q^p(x_q)\, {d\sigma_{DY}^{qN}(\tilde
x_1^q,M^2) \over d\tilde x_1^q}\over {d\sigma_{DY}^{pN}(M^2)\over dx_1} }.
\label{eq:alphx1} 
\end{eqnarray}

A distribution of energy losses occurs. We have calculated this 
distribution using Glauber theory. In the calculation, the
relatively small probability for inelastic collisions of the incident 
proton leads to a 
significant reduction of $\langle L\rangle=\langle z_2-z_1\rangle$ with 
respect to the mean path length $L_0$. For example, for tungsten we 
find $\langle L\rangle =2.4$ fm, whereas for a uniform sphere 
$L_0=3R_0A^{1/3}/4 =4.9$ fm.

Shadowing is the well-known 
reduction of the cross section per nucleon, observed
experimentally in DIS\cite{geesaman} for $x$ less than about 0.07.  
In proton-nucleus DY production a reduction in the cross section per nucleon 
at small $x$ is seen at the highest
available proton energy, 800 GeV/c\cite{aldedy,e866,fsl}. It is, however, an open 
question whether this is shadowing, energy loss, or both. The 
framework for analyzing shadowing in the rest frame of the nucleus, given 
below, is crucial to resolving this puzzle\cite{hir,kst1,kst2,krt2}.

\begin{figure}[thb]                                               
  \begin{center}
    \mbox{\epsfxsize=2.25in\epsffile{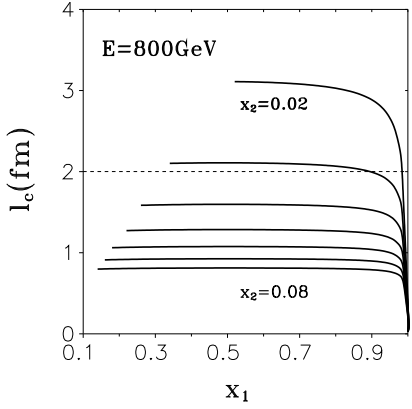}}
    \vspace*{-9pt}                            
  \end{center}     
\caption{The mean coherence length as function of $x_1$
at fixed values of $x_2=0.02,\ 0.03,\ ...\ 0.08$, evaluated for the
kinematic conditions of E772.}                                          
  \label{l_c}
\end{figure}

In both DIS and DY shadowing occurs when the nuclear coherence
length grows larger than the distance between nucleons, $\approx 2$ fm. 
The coherence length is a measure of the lifetime of the fluctuation 
of a quark into a virtual photon and residual quark. For the DY
process, the mean nuclear coherence length is given\cite{kst1,longpaper} by
\begin{equation} 
l_c=\left\langle\frac{2\,E_q\,x^q_1\,(1-x^q_1)}
{(1-x^q_1)\,M^2+(x^q_1\,m_q)^2+k_T^2}\right\rangle\ , 
\label{eq:l_c} 
\end{equation}
 where $E_q=x_q\,E_p$ and $m_q$ are the energy and mass of the projectile quark
which radiates the virtual photon.  The resulting lepton pair has an effective mass
$M$, a transverse momentum $k_T$, and carries a fraction $x^q_1$ of the
initial momentum of the quark.  The mean coherence length for
the kinematic conditions of E772 has been evaluated in Ref.\cite{longpaper} 
by integrating over $x^q_1$ and $k_T$. This is similar to a previous 
very successful treatment of shadowing in DIS\cite{krt2}. 
The result is shown versus $x_1$ in Fig.~\ref{l_c} for various fixed values 
of $x_2$.  
The coherence length is nearly independent of $M^2$ at fixed $x_2$, 
but it vanishes at $x_1\to 1$, violating factorization.
We note that the values of $l_c$ given by Eq.~\ref{eq:l_c} are 
significantly smaller than the commonly cited 
$l_c\approx 1/2m_Nx_2$ (see discussion in \cite{krt2}). 

In the weak shadowing 
approximation\cite{kp},  
\begin{equation}        
R^{shad}_{A/N}(x_1,M^2)\approx                            
1\,-\,{1\over4}\,\sigma_{\!e\!f\!f}\,\langle T_A\rangle\,    
F^2_A(q_c)\ .                                                            
\label{shadowing}                                                      
\end{equation}
This expression is
accurate for the whole kinematic range of E772.
Here, $\langle T_A\rangle$ is the mean value of the nuclear 
thickness function, $q_c=1/l_c$, and 
\begin{equation}
F^2_A(q_c) = \frac{1}{A\langle T_A\rangle}
\int d^2b\left|\int\limits_{-\infty}^{\infty}
dz\,e^{iq_cz}\,\rho_A(b,z)\right|^2\
\label{formfactor}
\end{equation}
is the longitudinal nuclear formfactor \cite{kk}, where
$\rho_A(b,z)$ is the nuclear density. 
The effective cross section is defined as \cite{hir}
$\sigma_{\!e\!f\!f}=\langle\sigma^2(x^q_1r_T)\rangle/
\langle\sigma(x^q_1r_T)\rangle$, where $\sigma(x^q_1r_T)$
is the $q\bar q$ dipole cross section \cite{zkl}, and $r_T$ is transverse 
separation between the virtual photon and the quark (Fig.~\ref{models}); 
$\sigma_{\!e\!f\!f}$ is in the range 3.5-5.5~mb for E772 \cite{longpaper}.

Energy loss and shadowing provide mechanisms for nuclear suppression in 
the DY process that are effective in different regimes.
If $l_c$ is short, no shadowing occurs
($F_A^2(q_c)\to 0$ in Eq.~\ref{shadowing}), but energy loss
can reduce the yield of DY pairs. In the opposite limit, $l_c\gg R_A$,
shadowing achieves its full strength ($F_A^2(q_c)\to 1$).
Here, initial state interactions 
do not affect the DY cross section except for transverse momentum 
broadening\cite{bbl,jkt}. Our ansatz is that the transition 
between these limiting regimes is controlled by $F_A^2(q_c)$.
The only expression which is linear in $F_A^2(q_c)$ and has the right
limits at $q_c\to 0$ and $q_c\to\infty$ reads
\begin{eqnarray} 
R_{A/N}(x_1,M^2)\,&=&\,
\Bigl(R^{\Delta E}_{A/N}(x_1)-1\Bigr)\,
\Bigl(1-F_A^2(q_c)\Bigr)\nonumber\\
&+&\,
R^{shad}_{A/N}(x_1,M^2)\ .
\label{suppression}
\end{eqnarray}

Given the above framework, the energy-loss analysis was accomplished as
follows. 
The quark-nucleon cross section in Eqs.~\ref{eq:signn}
and~\ref{eq:alphx1} was assumed to have the form 
\begin{eqnarray}       
{d\sigma^{qN}_{DY}(M^2)\over dx_1^q}=K(M^2)\times(1-x_1^q)^m, 
\label{pd}
\end{eqnarray} 
where $K$ and $m$ were determined from a fit to the p+$^2H$ data 
with Eq.~\ref{eq:signn}. It was found that $m$ did not change significantly 
with mass bin; thus one slope parameter was sufficient to characterize the full 
range of the p+$^2H$ data. Because the mass-dependent normalization (K) of the 
$q-N$ cross section occurs in both numerator and denominator 
of Eq.~\ref{eq:alphx1}, the energy loss term in 
Eq.~\ref{suppression} becomes independent of mass. 

\begin{figure}[thb]
\begin{center}
    \mbox{\epsfxsize=3.0in\epsffile{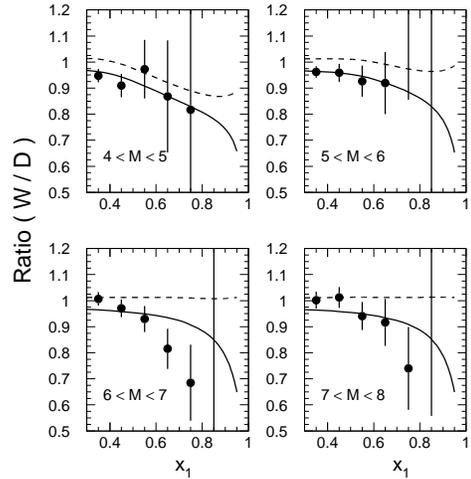}}
\vspace*{-1.8cm}
\end{center}
\caption{Ratio of tungsten to deuterium
Drell-Yan yields per nucleon versus $x_1$
 for different intervals of $M$.
Dashed curves correspond to net shadowing, 
solid curves show the full effect including 
shadowing and energy loss.}
  \label{w-d}
\end{figure}

\begin{figure}[thb]
\begin{center}
\vspace*{0.5in}
    \mbox{\epsfxsize=3.5in\epsffile{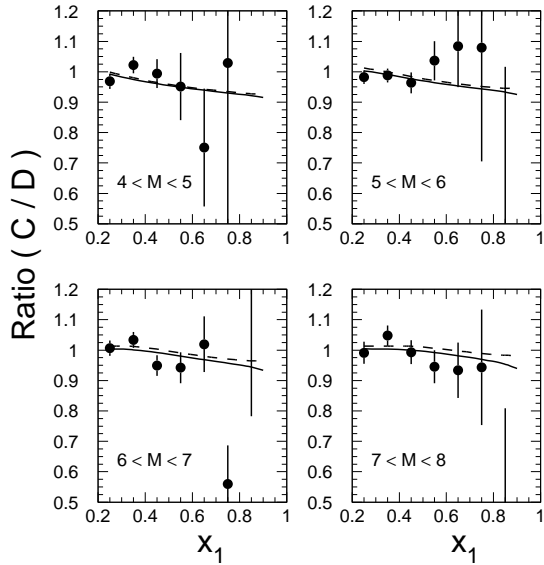}}
\vspace*{-2.8cm}
\end{center}
\caption{Ratio of Carbon to deuterium. Same labeling 
as Fig.~\ref{w-d}.}
  \label{c-d}
\end{figure}

Thus, in this formulation, energy loss and shadowing have different 
kinematic dependence. 
Therefore is essential to analyze nuclear-dependence ratios that are 
binned in dilepton mass. This allows $l_c$ to be evaluated for each bin 
in mass and $x_1$.
A two-parameter fit using Eq.~\ref{suppression} was applied to the 
heavy-target cross section ratios for C, Ca, Fe, and W; the parameters 
were $-dE/dz$ and an overall normalization factor, C.
The systematic normalization 
error, $\pm 1\%$, was treated as an additional statistical error.
The fit yields a substantial energy loss, 
$-dE/dz = 2.32\pm 0.52\pm 0.5$ GeV/fm (statistical and systematic), 
with $C = 1.010 \pm  0.006$, 
consistent with the E772 normalization uncertainty\cite{aldedy}. The 
systematic error associated with $-dE/dz$ arises 
from uncertainties in cut parameters, the range of applicability of 
Eq.~\ref{pd}, and the shadowing analysis (discussed below). 
Fits to $W/^2H$ and to $C/^2H$ in four mass intervals are shown by 
solid curves in Figs.~\ref{w-d} and~\ref{c-d}. 

\vspace*{1.0cm}

Our the analysis depends critically on
having separated the effects for energy loss and shadowing. 
Dashed curves show the net shadowing contribution. 
Nuclear suppression of the DY cross 
section for tungsten is mainly due to energy loss.
On the other hand, energy loss effects for carbon are small, 
the main contribution to nuclear suppression arising from shadowing.
This difference between the $A$- and $M$-dependence of energy loss and shadowing
permits the two effects to be disentangled.

We have checked the sensitivity of the results by the following tests:
(i) eliminated shadowing in Eq.~\ref{suppression} by
fixing $F_A^2=0$ ($-dE/dz =2.24 \pm 0.53$);
(ii) doubled the shadowing corrections, $1-R_{A/D}^{shad} 
\Rightarrow 2(1-R_{A/D}^{shad})$ ($-dE/dz =2.64 \pm 0.53$); 
(iii) mixed energy loss and shadowing effects differently
from Eq.~\ref{suppression},
$R_{A/D}=R_{A/D}^{\Delta E}\times R_{A/D}^{shad}$ ($-dE/dz =2.35 \pm 
0.53$: This is the E866 procedure\cite{e866}.); (iv) selected 
for analysis only data with small $x_2 < x_2^{max}$:
Within error bars $-dE/dz $ is constant for $0.3\geq x_2^{max} \geq 0.12$.
Thus $-dE/dz $ is subject to only small variation within the error bars.
These modifications all lead to a significant growth of $\chi^2$.

Recently the E866\cite{e866} collaboration analyzed DY nuclear 
dependence data from targets of Be, Fe, and Cu.  The E866 data set,
$1.3\times10^5$ muon pairs    
in the range $4\leq M\leq 8.5$ GeV/$c^2$, was more concentrated          
at low $x_2$ than the present one. E866 subtracted a 
phenomenological shadowing contribution, yielding new 
``shadowing-corrected'' nuclear dependence
ratios. Shadowing was calculated employing the
results of the global phenomenological analysis of DIS and DY data 
by Eskola et al.\cite{eskola} (EKS). The EKS analysis itself included the DY 
data
from E772, with the presumption that the low-$x_2$ nuclear dependence 
arose entirely from shadowing. This clearly introduced an inconsistency 
into the E866 search for energy loss. Considering the critical 
importance of separating shadowing and energy loss, which, at 800 GeV, 
can be achieved only via mass-binned nuclear-dependence ratios, it 
is not surprising that the E866 analysis missed the effect.

The value of $dE/dz$ determined here is not very different from that 
found many years ago by Gavin and Milana (GM)\cite{gm} using the 
mass-averaged W-D ratio 
from E772. In our model shadowing is a very small effect 
for the mass-averaged W data, so the GM analysis, which ignores shadowing, 
should 
not be too far off. The GM value, $dE/dz \approx$ 1.5 GeV/fm, should be 
increased by a factor of $\approx$ 2 to account for the reduction of the 
effective nuclear path length, discussed earlier in connection with 
Eq.~(\ref{eq:alphx1}). Unlike the GM model, our analysis presumes a 
constant $dE/dz$, yielding an energy loss that is 
independent of laboratory beam energy (see Ref.~\cite{bh}).

Much theoretical attention has been devoted in recent years to the 
elucidation of the QCD analogue of the famous 
Landau-Pomeranchuk-Migdal\cite{lp} 
effect(see Ref.~\cite{bsz}). Gluon 
radiation induced when a quark penetrates 
nuclear matter leads to additional energy loss 
proportional to the square of the path length traversed. Using the 
measured transverse-momentum broadening\cite{aldeups} this can be estimated for 
Tungsten as rising to a maximum value, $-(dE/dz)_{rad}\approx 0.2$ GeV/fm. Thus 
for cold matter radiative energy loss is not a large contribution to the 
total.

In light of the present finding that quark energy loss is significant, 
one should re-examine the role of energy loss in the nuclear dependence 
of \J 
production (see Ref.~\cite{mmp}). It was 
demonstrated many years\cite{kn1,katsanevas} ago that the $x_F$-dependence of 
\J suppression at energies $150-300$ GeV could be well described by energy 
loss, with $-dE/dz\approx$ 3-4 GeV/fm. The larger value for a 
projectile gluon is consistent with an enhancement due to the 
Casimir factor, $9/4$. 

In summary, we have made the first determination of quark energy loss 
using an analysis that takes into account nuclear shadowing.
The result, $-dE/dz = 2.32 \pm 0.52\pm 0.5$ 
GeV/fm, is in approximate accord with the theoretical expectation that 
energy loss should be at least the order of the string tension, 
$\kappa\approx$ 1 GeV/fm. At 800 GeV mass binning of the nuclear 
dependence ratios is 
crucial to the separation of energy loss and shadowing effects. 
It would 
be very desirable to have precise measurements of the nuclear 
dependence of DY production at lower beam energies (100-300 GeV) where 
shadowing disappears and energy loss would provide the dominant nuclear 
dependence.

\end{document}